# Collaborative Anomaly Detection Framework for handling Big Data of Cloud Computing

*Nour Moustafa, Gideon Creech, Elena Sitnikova, Marwa Keshk*
School of Engineering and Information Technology
University of New South Wales at the Australian Defence Force Academy, Canberra, Australia
E-mail: nour.abdelhameed@student, g.creech, e.sitnikova, marwa.hassan@student_@{.adfa.edu.au}

*Abstract*— **With the ubiquitous computing of providing services and applications at anywhere and anytime, cloud computing is the best option as it offers flexible and pay-per-use based services to its customers. Nevertheless, security and privacy are the main challenges to its success due to its dynamic and distributed architecture, resulting in generating big data that should be carefully analysed for detecting network's vulnerabilities. In this paper, we propose a Collaborative Anomaly Detection Framework (CADF) for detecting cyber attacks from cloud computing environments. We provide the technical functions and deployment of the framework to illustrate its methodology of implementation and installation. The framework is evaluated on the UNSW-NB15 dataset to check its credibility while deploying it in cloud computing environments. The experimental results showed that this framework can easily handle large-scale systems as its implementation requires only estimating statistical measures from network observations. Moreover, the evaluation performance of the framework outperforms three state-of-the-art techniques in terms of false positive rate and detection rate.**

## I. INTRODUCTION

The term 'cloud computing' denotes a network of networks interconnected using internet services in which virtual shared servers offer the software, infrastructure, platform, services and other resources to customers anywhere and anytime [1]. Cloud computing produces a flexible computing model which permits firms and organisations to use and adapt their IT needs over the internet at a low cost of use and without any liability towards IT infrastructure and maintenance [2].

In the cloud computing environment, network-accessible resources are used as services. These services are categorised into three types of Platform as a Service (PaaS), Infrastructure as a Service (IaaS) and Software as a Service (SaaS) models [2] [3] [4]. Firstly, a PaaS delivers to a user or organisation client applications using programming languages, libraries, services and tools which are supported by a PaaS provider's infrastructure. Then, an IaaS offers processing units, network capabilities and other fundamental computing resources via Virtual Machines (VMs) to service subscribers. Finally, a SaaS offers to a user or organisation on-demand applications and software services via a cloud infrastructure, avoiding the cost of buying and maintaining those applications.

Cloud executions often contain security mechanisms, typically available because of the data centralisation and global architecture. Cloud providers endeavour to secure the homogeneous resources of cloud architecture as much as possible [3]. However, several vulnerabilities are a result of the underlying technologies, for example, network systems, APIs, datacentres and virtual machines that considerably threaten the cloud architecture [2].

The architecture of cloud computing includes three layers: infrastructure, application and platform which execute its functionalities. Each layer faces particular vulnerabilities, developed by diverse malicious scripts or configuration errors of user/service providers. Cloud's vulnerabilities expose the confidentiality, integrity or/and availability of its resources. This is because that data and virtualised infrastructure of cloud systems can be breached by existing and new attacks [5]. The security challenge of a cloud computing system occurs when a cloud runs a high storage capacity and computing power that is abused by an insider or outsider hacker [6].

There are some existing security techniques and tools, including authentication, access control, encryption, access control, firewall and intrusion detection systems (IDSs), to tackle the cloud's security issues. However, in current cloud computing systems, no single mechanism fits all cases of exploitation. These mechanisms should be incorporated to produce a comprehensive layer of defence. In this study, we mainly focus on the IDS technology and what is the suitable framework for detecting intrusive events that threaten cloud environments.

We propose a Collaborative Anomaly Detection Framework (CADF) for processing big data of cloud computing systems. More specifically, we provide the technical functions and the way of deployment of this proposed framework for these environments. The technical framework comprises three modules: capturing and logging network data, pre-processing these data and a new Decision Engine (DE) using a Gaussian Mixture Model (GMM) [15] and lower-upper Interquartile Range (IQR) threshold [16] for detecting attacks. The UNSW-NB15 dataset[1] is used for evaluating the new DE to assess its reliability while deploying the framework in real cloud computing systems.

## II. BACKGROUND AND RELATED WORK

Because of the dynamic configurations of cloud computing, numerous vulnerabilities attempt to penetrate its architecture, leaving loopholes in which attackers exploit cloud's services and its big data [2]. The analysis of cloud data should consider the inspection of big data properties, i.e., *volume*, *velocity*, *variety, veracity* and *value,* for efficiently detecting malicious activities [7]. Inspecting these properties in cloud data helps in making the decision of designing a scalable security mechanism that can precisely model network data for defining malicious patterns, and these properties are declared as follows.

- *Volume* is a large amount of processed data.
- *Velocity* is the high speed of processed data.
- *Variety* is the dimensionality of processed data.
- *Veracity* is the correctness of processed data.
- *Value* is the significance of processed data.

An IDS is widely used to detect intrusive activities from cloud's big data, but it still faces the challenge of successfully recognising invariants of known attacks and zero-day/new malicious activities. The purpose of IDS is to provide a layer of defence against malicious events that try to breach computing systems. It monitors and analyses activities which happen in computer or network systems to detect possible threats [1].

The IDS detection approaches are classified into three categories: misuse- (MDS), anomaly- (ADS) and hybrid-based IDS, merging the first two types [1][5][7]. A MDS monitors network data to match observed activities against an existing blacklist. Nevertheless, although it produces high detection rates and low false positive rates, it cannot identify any zero-day attacks or even variants of known ones [5]. Conversely, an ADS establishes a normal profile and discovers any variation from it as an attack. Because it can identify both known and unknown attacks, it is a better approach than a MDS if its detection method is properly developed [1] [5] [3].

The majority of recent cloud computing IDS research focuses on its design at the application, platform, and infrastructure layers separately [6]. For instance, Gustavo and Miguel [8] executed many ADS techniques and suggested an IDS for protecting complex web applications as SaaS. Their results showed that the deployment of ADS at the application layer is very effective, as it is easy to detect application attacks. Nevertheless, they did not provide an effective way of deploying their system in a real cloud computing environment.

Establishing the IDS in the infrastructure layer is important to some extent. As in [9], the authors suggested a hypervisor model based on a VM monitor to secure the infrastructure layer (IaaS) from different types of attacks. This model enhances the reliability and availability of the system because the running services can be protected. However, this model cannot protect the system if the infrastructure collapses due to the norm of contemporary flooding traffic of attacks such as DDoS.

Designing a collective IDS structure for cloud computing systems is always an arduous task because of their heterogeneous model and virtualisation technology. Zayed et al. [10] developed a collaborative IDS using a support vector machine technique for detecting abnormal activities. However, this system is not scalable as the performance drops with the increase of data capacity into the central node in which a single point of failure is unsuitable in the cloud.

Gai et al. [11] suggested a grid and cloud computing IDS for discovering malicious events. However, this system can only detect particular attacks. Tan et al. [2] proposed a collective IDS which associates malicious events between different IDSs to enhance the IDS efficiency. Although these collaborative systems are scalable to some extent, they cannot efficiently detect large-scale distributed anomalies, and there is no central correlation handler to merge activities, as we propose in this study.

## III. PROPOSED COLLABORATIVE ANOMALY DETECTION FRAMEWORK (CADF)

Existing misuse IDSs are not able to identify zero-day attacks or even variants of existing types. The design of a collaborative IDS for each node in cloud computing environments is extremely significant for detecting these types of intrusions. A Collaborative Anomaly Detection Framework (CADF) is proposed to detect malicious observations from each network node in order to considerably improve the detection accuracy.

---

[1]The UNSW-NB15 dataset, https://www.unsw.adfa.edu.au/australian-centre-for-cyber-security/cybersecurity/ADFA-NB15-Datasets/ , May 2015

The target of the framework is to develop an effective ADS installed in each node in cloud computing systems which identifies malicious activities with a central data capturing and logging module. We describe the technical functions and deployment of this framework to understand the way of implementing it in real environments. The technical framework involves three modules, capturing and logging, data pre-processing and decision engine for identifying suspicious activities of cloud, as depicted in Figure 1.

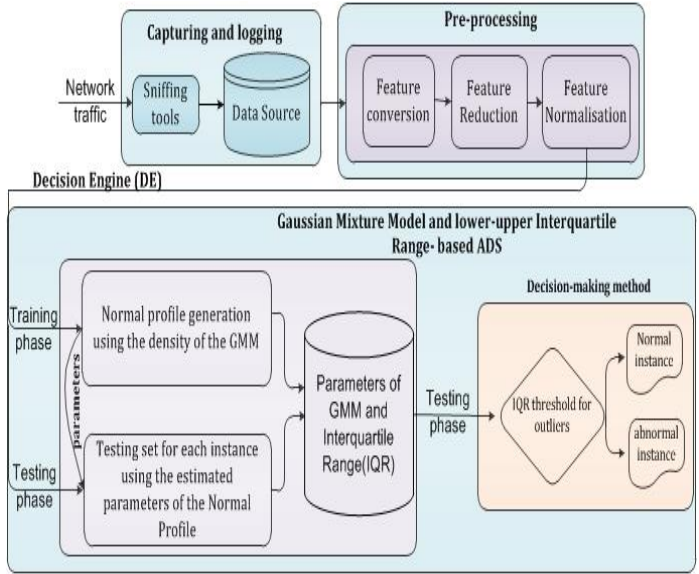

Figure 1. Proposed Collaborative Anomaly Detection Framework (CADF)

### A. Capturing and logging module

This module includes the steps of sniffing network data and storing them to be processed by the DE technique, like the steps of designing the UNSW-NB15 dataset [12] [13]. The configuration of the UNSW-NB15 testbed was used to simulate a large-scale network. A tcpdump tool was applied to sniff packets from the network's interface while Bro, Argus tools and other scripts were used to extract a set of features from network flows 12] [13]. These features were recorded using the MySQL Cluster CGE technology[2] that has a highly scalable and real-time database and enables a distributed architecture to read and write intensive workloads, and is accessed via SQL APIs for processing big data.

### B. Pre-processing module

This module determines and filters network data in three steps. Firstly, its feature conversion replaces non-numeric features with numeric ones because our GMM-based ADS technique deals with only numeric features, for example mapping TCP, UDP and ICMP into 1, 2 and 3, respectively. Secondly, its feature reduction uses the PCA technique to select a small number of uncorrelated features. Because this is one of the best-known linear feature reduction algorithms, with the advantages of demanding less memory storage, having lower data transfer and processing times, and better accuracy than others [14], we used it for this study.

Finally, feature normalisation arranges the value of each feature in a certain range to remove any bias from raw data and easily process it. We apply the z-score function, as it can scale the network data with no change in the norm of the original data. It scales each feature ($x$) with a 0 mean ($\mu$) and 1 standard deviation ($\delta$), to normalise the data using (1).

$$z = \frac{x-\mu}{\delta} \qquad (1)$$

## IV. DECISION ENGINE MODULE

This section elaborates the new DE technique based on the Gaussian mixture model and lower-upper interquartile range baseline.

### A. Finite Mixture Model using Gaussian distribution

As a finite mixture model is defined as a convex combination of two or more Probability Density Functions (PDFs), the joint properties of these functions can approximate any arbitrary distribution. It is a powerful and flexible probabilistic modeling technique for multivariate data [15]. Network data are typically considered multivariate as they have $d$ dimensions for differentiating between attack and normal instances; for example, let $X = [X_1, \ldots, X_d]$ be a d-dimensional random variable and $x = [x_1, \ldots, x_d]$ an observation of $X$. The probability density Function (PDF) of a Gaussian distribution is computed by

$$f(x|\mu, \delta^2) = \frac{2}{\sqrt{2\pi\delta^2}} e^{-\frac{(x-\mu)^2}{2\delta^2}} \qquad (2)$$

where $x$ is feature values, $\mu$ is mean of the distribution and $\delta^2$ is variance. The PDF of a mixture model is declared by a convex combination of $K$-component PDFs and is given as

$$p(x|\theta) = \sum_{k=1}^{K} \alpha_k\, p(x|\theta_k) \qquad (3)$$

where $(\alpha_1, \ldots, \alpha_k)$ are the mixing proportions, each $\theta_k$ is a set of parameters defining the $k$ components which are based on the number of the feature selected using the PCA technique and $\theta = (\theta_1, \ldots, \theta_k, \alpha_1, \ldots, \alpha_k)$ is the complete set of parameters required to identify the mixture. Applying the probability conditions, $\alpha_k$ has to satisfy

$$\alpha_k \geq 0,\, K = 1, \ldots, k \text{ and } \sum_{k=1}^{K}\alpha_k = 1 \qquad (4)$$

---

[2] The MySQL CGE technology, https://www.mysql.com/products/cluster/ , May 2017.

The mixture model is computed by the Maximum Likelihood Estimation (MLE) [15]. Assuming $X$ data with $N$ observations, the probability of data in which $x_i$ are identically and independently distributed is given by

$$p(X|\theta) = \mathcal{L}(\theta|X) = \prod_{i=1}^{N} \sum_{k=1}^{K} \alpha_k \, p_k(x_i|\theta_k) \quad (5)$$

The MLE is derived from the set of parameters ($\theta$) by

$$\theta^* = argmax_\theta \, \mathcal{L}(\theta|X) \quad (6)$$

The GMM is the mixture model most often applied for NADSs. It estimates the PDFs (from equations (2) to (6)) of the normal data given by a training set. The parameters $\theta = (\alpha, \mu, \delta)$ of the GMM are estimated using the EM algorithm to model network data.

### B. Training Phase

It is vital to obtain a purely normal training set to assert correct detection. Given a set of normal vectors ($r_{1:n}^{normal}$) in which each record comprises a set of features, where $r_{1:n}^{normal} = \{x_1, x_2, ..., x_D\}^{normal}$, the normal profile contains only statistical measures from $r_{1:n}^{normal}$. They involve the estimated parameters $\theta = (\alpha, \mu, \delta)$ of the GMM to compute the PDFs of the Gaussian distribution ($GMM(X|\alpha, \mu, \delta)$) for each vector in the training set, as shown in Figure 2.

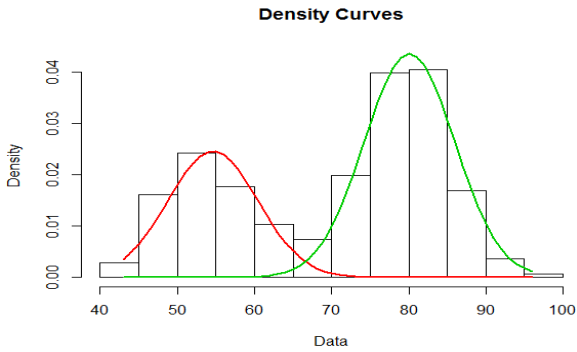

Figure 2. Sample of density curves for normal network data

Algorithm 1 presents the proposed steps for establishing a normal profile (pro), with the parameters ($\alpha, \mu, \delta$) of the GMM estimated for all the normal vectors ($r_{1:n}^{normal}$) using the equations published in [15], and then the PDFs of the features ($x_{1:D}$) are calculated using equations (2) to (6). Following this, the IQR is calculated by subtracting the first quartile subtracted from the third quartile of the PDFs [16] to generate a threshold for identifying abnormal instances in the testing phase. It is known that quartiles divide data into contiguous intervals with equal probabilities [16].

Algorithm 1: generation of normal profile in training phase

**Input:** normal vectors ($r_{1:n}^{normal}$)
**Output:** normal profile (pro)
1. **for each** record i in ($r_{1:n}^{normal}$) **do**
2. estimate the parameters ($\theta = (\alpha, \mu, \delta)$) of the GMM
3. compute the PDFs using equations (2) to (6) based on the parameters estimated in Step 2
4. **end for**
5. calculate lower = quartile (PDFs,1)
6. calculate upper = quartile (PDFs,3)
7. calculate IQR = upper - lower
8. pro ← {($\theta = (\alpha, \mu, \delta)$), (lower-upper IQR)}
9. **return** pro

### C. Testing Phase

In the testing phase, the Gaussian PDF ($PDF^{testing}$) of each vector ($r^{testing}$) is calculated using the same parameters computed for the normal profile (pro). Algorithm 2 describes the steps in the testing phase and decision-making method for specifying the Gaussian PDFs of attack records, with step 1 building the PDF of each vector using the stored normal parameters ($\theta = (\alpha, \mu, \delta)$).

Algorithm 2: testing phase and decision-making method

**Input:** observed record ($r^{testing}$)
, normal profile (pro) {($\theta = (\alpha, \mu, \delta)$), (lower-upper IQR)}
**Output:** normal or abnormal record
1. compute the $PDF^{testing}$ using equations 2 to 6 with parameters ($\theta = (\alpha, \mu, \delta)$)
2. **if** ($PDF^{testing}$ < (lower –w.(IQR))) || ($PDF^{testing}$ > (upper + w.(IQR)) **then**
3. **return** attack
4. **else**
5. **return** normal
6. **end if**

Steps 2 to 6 are the steps of the decision-making method. The IQR of the normal vectors is calculated to find the anomalies of any testing record ($r^{testing}$) which are considered to be any vector falling below *(lower – w.(IQR))* or above *(upper + w.(IQR))*, where w is interval values between 1.5 and 3 that precisely represents the lower and upper bounds of normal data, as proven in [16]. The detection decision is based on considering any $PDF^{testing}$ falling outside of this interval as anomalies, otherwise they are normal records.

## V. DEPLOYMENT OF PROPOSED FRAMEWORK FOR CLOUD COMPUTING ENVIRONMENTS

The deployment of this framework is described for three nodes (A, B and C) depicted in Figure 3 in order to be executed for cloud computing systems. Unlike traditional IDSs, the CADF is deployed on each network node and each CADF connected simultaneously with the shared module of capturing and logging. This is for collecting attribute values of network traffic in a particular time interval to make it much easier while passing the processed data to the DE module for each network node.

We suggest the deployment in two stages: a shared module as SaaS and ADS as SaaS. The first includes a sensor for capturing network attributes and logging them in a data source, as presented in Figure 3. It is designed to be a sharable service for the entire connected ADSs at different cloud nodes. The second contains the main functionality of the proposed ADS to be installed at each node for handling large-scale networks by distributing it as service at each node.

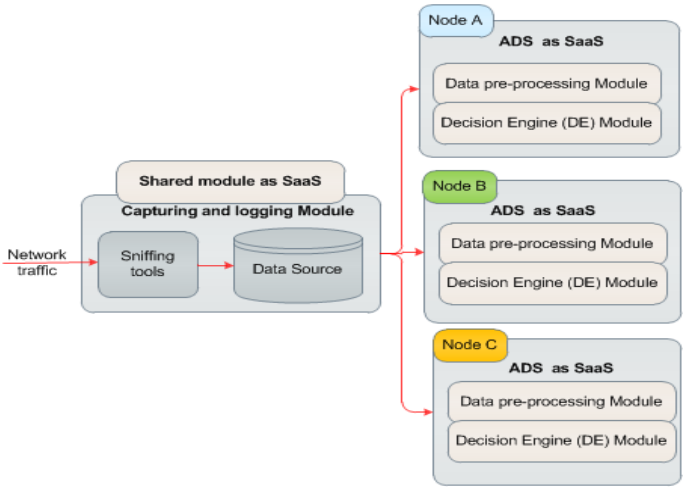

Figure 3. Deployment architecture of CADF

## VI. EXPERIMENTAL RESULTS AND EXPLANATIONS

### A. Dataset and pre-prosessing module used for evaluation

The evaluation of the proposed framework is conducted using the UNSW-NB15 dataset which has a hybrid of authentic contemporary normal and attack vectors. The volume of its network packets is nearly 100 Gigabytes, generating 2,540,044 observations which are recorded in four CSV files. Each record includes 47 features and the class label. The dataset comprises ten different classes, one normal and nine types of security events and malware (i.e., Analysis, Backdoors, DoS, Exploits, Generic, Reconnaissance, Fuzzers for anomalous activity, Shellcode and Worms). The GMM-ADS technique is evaluated using 10 features selected from the UNSW-NB15 dataset selected using the PCA, as presented in Table I.

Table I. FEATURES SELECTED FROM UNSW-NB15 DATASET

*ct_dst_sport_ltm, tcprtt, dwin, ct_src_dport_ltm, ct_dst_src_ltm, ct_dst_ltm, smean, dmean, service, proto*

The proposed technique was developed using the 'R language' on Linux Ubuntu 14.04 with 16 GB RAM and an i7 CPU processor. To conduct the experiments on the dataset, we selected random samples from the CSV files of the UNSW-NB15 dataset with various sample sizes between 70,000 and 150,000. In each sample, normal records were about 60-70% of the total size, with some used for establishing the normal profile and the testing set.

### B. Performance Evaluation

The accuracy, Detection Rate (DR) and False Positive Rate (FPR) explained below are used to evaluate the framework performance.

- The **accuracy** is the percentage of all normal and attack records correctly classified, that is,
$$Accuracy = \frac{TP+TN}{TP+TN+FP+FN} \quad (7)$$
- The **DR** is the percentage of correctly detected attack records, that is,
$$Detection\ Rate = \frac{TP}{TP+FN} \quad (8)$$
- The **FPR** is the percentage of incorrectly detected attack records, that is,
$$False\ Positive = \frac{FP}{FP+TN} \quad (9)$$

where TP (true positive) is the number of actual attack records classified as attacks, TN (true negative) is the number of actual normal records classified as normal, FN (false negative) is the number of actual attack records classified as normal and FP (false positive) is the number of actual normal records classified as attacks.

### C. Result discussion

The performance evaluation of the CADF was conducted on the features selected from the UNSW-NB15 dataset, with the overall DR, accuracy and FPR values demonstrated in Table II. Figure 4 presents the Receiver Operating Characteristics (ROC) curves which display the relationship between the DRs and FPRs using the w values.

Table II. Evaluation of features from unsw-NB15 dataset

| w value | DR | Accuracy | FPR |
|---|---|---|---|
| 1.5 | 86.3% | 88.2% | 8.4% |
| 2 | 89.1% | 90.1% | 5.5% |
| 2.5 | 93.4% | 94.8% | 4.4% |
| 3 | 95.6% | 96.7% | 3.5% |

It can be seen that the stable increase in the w value between 1.5 and 3 increased the overall DR and accuracy while decreasing the overall FPR. The overall DR and accuracy increased from 86.3% to 95.6 % and 88.2% and 96.7%, respectively, however the overall FPR decreased from 8.4 % to 3.5% when the w value increased from 1.5 to 3.

The key reasons for the CADF performing better than the other peer techniques discussed below are that the GMM can perfectly fit the boundaries of each feature as it accurately estimates the mixing weights of network features in order to model normal data. Moreover, the lower-upper IQR method can successfully specify the boundaries between normal and outlier observations.

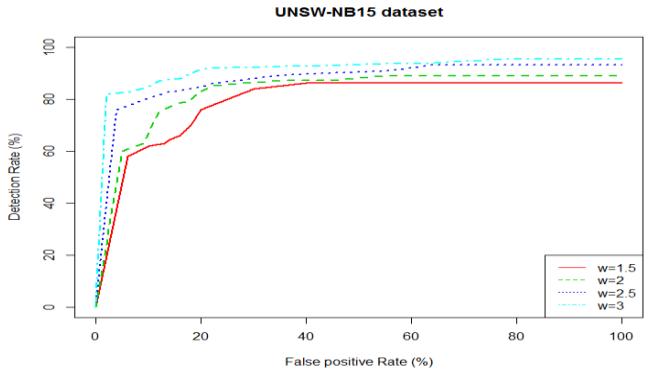

Figure 4. ROC curves with w values

The performance evaluation results for the CADF were compared with three existing techniques, namely the Triangle Area Nearest Neighbours (TANN) [17], Euclidean Distance Map (EDM) [18] and Multivariate Correlation Analysis (MCA) [19], with their overall DRs and FPRs listed in Table III, revealing the superiority of our framework.

Table III. Comparison of performances of four techniques

| Technique | DR | FPR |
|---|---|---|
| TANN [17] | 88.2% | 12.3% |
| EDM [18] | 89.4% | 10.6% |
| MCA [19] | 91.4% | 8.9% |
| Proposed CADF | 95.6% | 3.5 % |

According to the above discussions, the proposed framework can be easily deployed in cloud computing systems. Since the shared module as SaaS collects important network observations from different network nodes, the DE as SaaS does not consume high processing time to inspect the observations, either normal or attacks, for these nodes. This is because the DE was built based on only estimating statistical measures of Gaussian mixture model and lower-upper interquartile range from network instances that can be simply computed in real cloud computing systems with less computational resources.

## VII. CONCLUDING REMARKS

This study discusses a new collaborative anomaly detection framework for detecting known and unknown intrusive activities in cloud computing environments. This framework comprises capturing and logging network data, pre-processing these data to be handled at the decision engine sensor and a new decision engine using the Gaussian Mixture model and interquartile range for identifying abnormal patterns. Moreover, the architecture for deploying this framework as Software as a Service (SaaS) is produced in order to be easily installed in cloud computing systems. The experimental results of the framework show its superiority for detecting abnormal events using the UNSW-NB15 dataset compared with three ADS techniques. In future, we plan to extend this study for deploying the framework in a real cloud computing environment with further findings and explanations.

REFERENCES

1. Shelke, Ms Parag K., Ms Sneha Sontakke, and A. D. Gawande. "Intrusion detection system for cloud computing." *International Journal of Scientific & Technology Research* 1.4 (2012): 67-71.
2. Tan, Zhiyuan, et al. "Enhancing big data security with collaborative intrusion detection." *IEEE cloud computing* 1.3 (2014): 27-33.
3. Zissis, Dimitrios, and Dimitrios Lekkas. "Addressing cloud computing security issues." *Future Generation computer systems* 28.3 (2012): 583-592.
4. Somani, Gaurav, et al. "DDoS attacks in cloud computing: issues, taxonomy, and future directions." Computer Communications (2017).
5. Samaila, Musa G., et al. "Security Challenges of the Internet of Things." Beyond the Internet of Things. Springer International Publishing, 2017. 53-82.
6. Patel, Ahmed, et al. "An intrusion detection and prevention system in cloud computing: A systematic review." *Journal of* network and computer applications 36.1 (2013): 25-41.
7. Zuech, Richard, Taghi M. Khoshgoftaar, and Randall Wald. "Intrusion detection and big heterogeneous data: a survey." Journal of Big Data 2.1 (2015): 3.
8. Nascimento, Gustavo, and Miguel Correia. "Anomaly-based intrusion detection in software as a service." Dependable Systems and Networks Workshops (DSN-W), 2011 IEEE/IFIP 41st International Conference on. IEEE, 2011.
9. Iqbal, Salman, et al. "On cloud security attacks: A taxonomy and intrusion detection and prevention as a service." Journal of Network and Computer Applications 74 (2016): 98-120.
10. Al Haddad, Zayed, Mostafa Hanoune, and Abdelaziz Mamouni. "A Collaborative Network Intrusion Detection System (C-NIDS) in Cloud Computing." International Journal of Communication Networks and Information Security 8.3 (2016): 130.
11. Gai, Keke, et al. "Intrusion detection techniques for mobile cloud computing in heterogeneous 5G." Security and Communication Networks 9.16 (2016): 3049-3058.
12. Moustafa, Nour, and Jill Slay. "UNSW-NB15: a comprehensive data set for network intrusion detection systems (UNSW-NB15 network data set)." Military Communications and Information Systems Conference (MilCIS), 2015. IEEE, 2015.
13. Moustafa, Nour, and Jill Slay. "The evaluation of Network Anomaly Detection Systems: Statistical analysis of the UNSW-NB15 data set and the comparison with the KDD99 data set." Information Security Journal: A Global Perspective 25.1-3 (2016): 18-31.
14. Vasan, K. Keerthi, and B. Surendiran. "Dimensionality reduction using Principal Component Analysis for network intrusion detection." Perspectives in Science 8 (2016): 510-512.
15. Gelman, Andrew, et al. Bayesian data analysis. Vol. 2. Boca Raton, FL, USA: Chapman & Hall/CRC, 2014.
16. McLachlan, Geoffrey, and David Peel. Finite mixture models. John Wiley & Sons, 2004.
17. C. F. Tsai and C. Y. Lin, "A Triangle Area Based Nearest Neighbors Approach to Intrusion Detection," Pattern Recognition, vol. 43, pp. 222-229, 2010.
18. Z. Tan, A. Jamdagni, X. He, P. Nanda, and R. P. Liu, "Denial of-Service Attack Detection Based on Multivariate Correlation Analysis," Neural Information Processing, 2011, pp. 756-765.
19. Tan, Zhiyuan, et al. "A system for denial-of-service attack detection based on multivariate correlation analysis." Parallel and Distributed Systems, IEEE Transactions on 25.2 (2014): 447-456.